\documentclass[11pt,twoside]{article}

\usepackage{galev06}
\usepackage{epsf}
\usepackage{psfig}
\usepackage{lscape}
\usepackage{graphics}
\begin{document}
\title{Deep $GMOS$ Spectroscopy of Faint $SWIRE$ Populations.}   
\author
{M. Trichas$^1$,
M. Rowan-Robinson$^1$,
I. Waddington$^2$,
T.S.R. Babbedge$^1$}
\affil{$^1$Astrophysics Group, Blackett Laboratory, Imperial College London, Prince Consort Road, London, SW7 2BW, UK.\\
$^{2}$Astronomy Centre, Department of Physics \& Astronomy, 
        University of Sussex, Brighton, BN1 9QH, UK.}    

\begin{abstract} 
We have carried out a deep (21.5 $<$ R $< $23.9) systematic survey of galaxies selected from $SWIRE$ using $GMOS$ to target X-ray sources, AGN and galaxies with photometric redshifts greater than 1. We have obtained a total of 198 redshifts, of which 132 have secure redshifts in the range 0.12 $< $z $<$1.21. This sample, combined with the 111 redshifts targeted with $WIYN$ in the same field and the 1 square degree survey of $ELAIS-N1$ with $ACIS-I$, will allow us to calibrate our photometric redshift techniques; characterise the star formation history of the universe between 0.5 $<$ z$ <$ 1.2 where galaxy formation activity peaks; and explore the connection between $AGN$ and star-formation activity.
\end{abstract}


\section{Introduction}   
In recent years, extragalactic surveys have revolutionised our view of the high-redshift Universe, showing that the redshift range 0.5$<$z$<$3.0 witnessed an extraordinary transformation in the demographics of starburst and AGN activity in galaxies. The factors that drove these changes are however unclear. Consequently, the properties of large samples of sources at z$>$1 are the current driving force behind modern extragalactic surveys. At the forefront of these investigations is the SWIRE survey (Lonsdale et al. 2003). SWIRE has detected over 2 million infrared galaxies with the science objective of tracing the evolution of galaxies in the range 0.2$<$z$<$3 as a function of redshift, environment and luminosity. 
\section{Observations}  
 \begin{figure}
  \plottwo{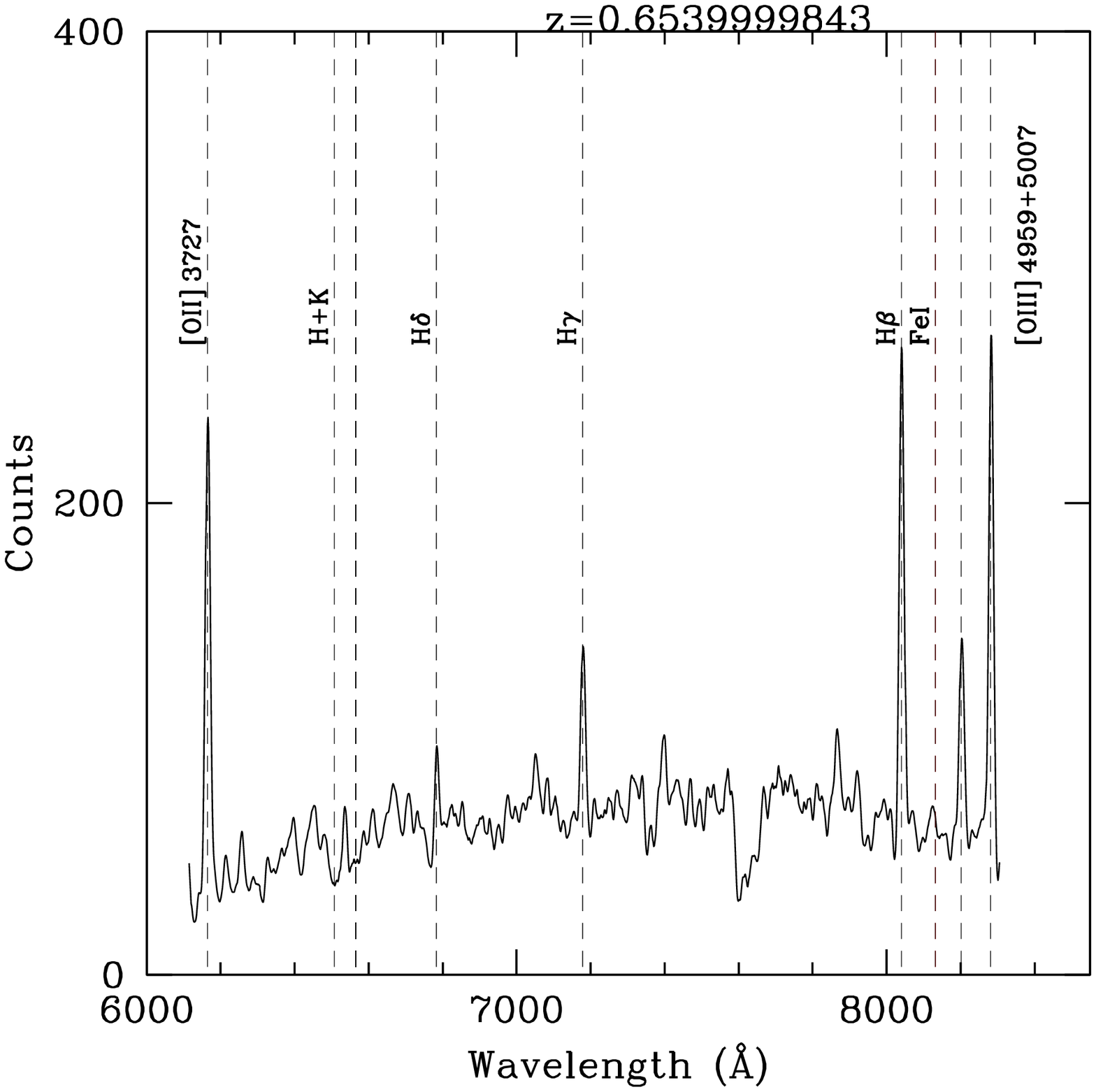}{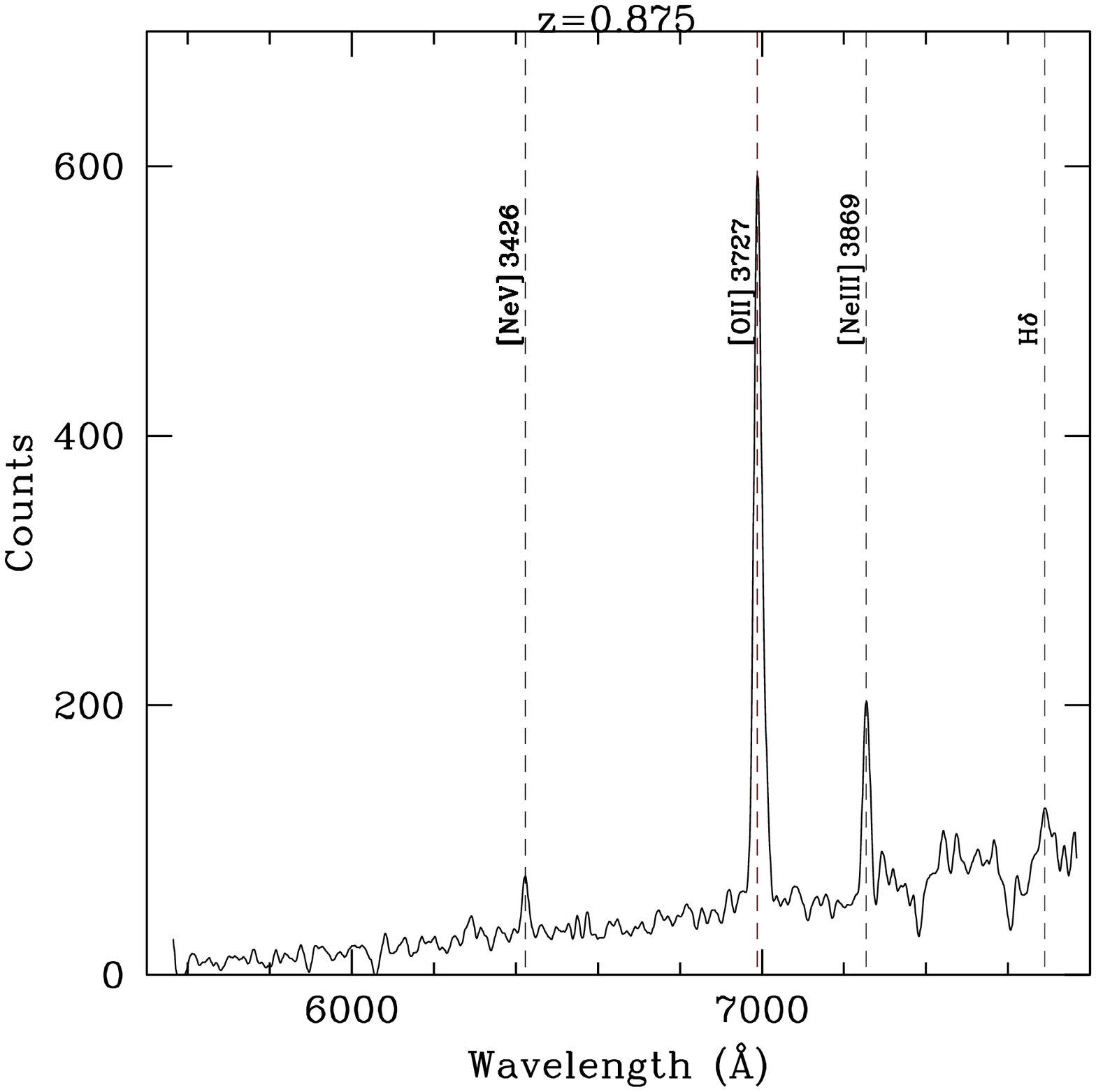}
   \caption{Representative spectra of two star forming galaxies.}
 \end{figure}
 We have used GMOS to target a total of 236 sources in 13 masks between $21.5<r<23.9$, 6 of them in the area covered by the ELAIS Deep X-ray Survey (Manners et al 2003) observing a total of 14 X--ray AGN,  one pointing within the 14~square arcmin ACS image and a further six
pointings, randomly scattered across the SWIRE field. All the observations were taken with the R400\_G5305  grating at a central wavelength of 7500~\AA whilst Nod and Shuffle was employed.A target list for each field was prepared from the photometric
redshift catalogue, selecting sources with $21.5<r<23.5$.  These
sources were then distributed between three prioritized groups, with
top priority given to X-ray sources in the {\it Chandra\/} fields.
Next priority were sources with either (i) optical or infrared SEDs
identified as AGN in the catalogue, or (ii) photometric redshifts
$z>1$.  Finally, all other SWIRE sources with $21.5<r<23.5$ were
assigned to the lowest priority group. The presence of strong emission and absorption lines ($[OII]$$\lambda$$3727$, $[OIII]$$\lambda$$4959+5007$, $H$$\alpha$, $H$$\beta$, $H$$\gamma$, $H$$\delta$, $H+K$, $MgII$)   immediately indicated the approximate redshift in 132 out of 198 extracted sources. In the case of the 42 single line objects the redshift was assigned to $[OII]$$\lambda$$3727$.
\section{Comparison with Photometric Redshifts and Discussion}   
The photometric redshifts are estimated using the updated version of ImpZ code (Babbedge et al 2004) including improvements based on studies in Rowan Robinson et al 2005.Ê One important change to the ImpZ implementation to the SWIRE data is to carry out a double pass of the code on the data, in order to deal with AGN dust tori successfully. Mid--IR excess is fit separately by more appropriate SEDs (Rowan-Robinson et al. 2006).\\
With a total of 198 reasonably secure spectra from GMOS and a further 111 from WIYN, we have performed one of the largest spectroscopic follow ups in ELAIS--N1. This sample combined with our large set of infrared data from SWIRE and X-ray data from the Shallow survey (Nandra 2004) will enable us to meet our scientific goals. 


\acknowledgements A detailed analysis of these follow--ups will be given in Trichas et al 2006 (in preparation).



\begin{thebibliography}{}
\bibitem{} Babbedge T., et al., 2004, MNRAS, 353, 654
\bibitem{} Lonsdale C., et al., 2003, PASP 115, 897
\bibitem{} Manners J., et al., 2003, MNRAS, 343, 293
\bibitem{} Nandra K., 2004, cxo..prop. 1799N
\bibitem{} Rowan-Robinson M., et al., 2005, AJ, 129, 1183
\bibitem{} Rowan-Robinson et al, 2006, astro-ph/0603737
\end{thebibliography}
\end{document}